\documentclass{PoS}
\usepackage{dsfont}

\title{Density of states techniques for fermion worldlines}

\ShortTitle{Density of states techniques for fermion worldlines}

\author{Christof Gattringer \footnote{Currently on leave of absence from Institute of Physics, University of Graz, 8010 Graz, Austria.} 
\vskip2mm

FWF - Austrian Science Fund, 1090 Vienna, Austria 

\vskip2mm
        
E-mail: \email{christof.gattringer@fwf.ac.at}   }

\abstract{Worldline representations were established as a powerful tool for studying bosonic lattice field theories at finite density. For fermions, however, the 
worldlines still may carry signs that originate from the Dirac algebra and from the Grassmann nature of the fermion fields. We show that a density of states 
approach can be set up to deal with this remaining sign problem, where finite density is implemented in a canonical approach by working with a fixed winding 
number of the fermion worldlines. We discuss the approach in detail and show first results of a numerical implementation in 2 dimensions.}

\FullConference{39th International Symposium on Lattice Field Theory - Lattice2022\\
		August 8 -  13, 2022\\
		Bonn, Germany}

\begin{document}

\section{Introductory comments}

\noindent
Essentially all lattice field theories allow for representations where the matter degrees of freedom are represented by worldlines and the gauge degrees 
of freedom by worldsheets (see, e.g., 
\cite{Endres:2006zh} -- \cite{Gattringer:2018ous} 
for some key examples). In several cases these representations completely remove complex action
problems that originate from finite chemical potential or a topological term 
\cite{Gattringer:2018dlw} -- \cite{Anosova:2022cjm}. However, if the matter is fermionic then the corresponding 
worldlines still carry signs that emerge from the Grassmann nature of the fields as well as from the $\gamma$-algebra. Thus Monte Carlo simulations of
worldlines were so far restricted to bosonic theories, with the exception of massless staggered fermions in 2d 
\cite{Gattringer:2015nea,Goschl:2017kml,Hirtler:2022ycl}, where the signs are known to be absent.

However, worldlines are a very powerful and conceptually elegant framework, e.g., the net particle number has the form of a topological invariant, such that 
it is worth exploring approaches that may overcome the remaining sign problem of fermionic worldlines. Here we present a first exploratory
study to address this sign problem with a suitable density of states (DoS) approach (see also \cite{Francesconi:2021suq} for experiments in this direction). DoS techniques were
initially introduced to lattice field theory in \cite{Gocksch1,Gocksch2}, but saw a major revival  based on a paper by Langfeld, Lucini and Rago \cite{Langfeld:2012ah},
where a modern formulation with considerably improved accuracy was presented (see also 
\cite{Langfeld:2013xbf} -- \cite{Francesconi:2019nph}). 
Based on a variant 
\cite{Z3_FFA_2} -- \cite{Gattringer:2021xrb} 
of these developments, the functional fit approach (FFA), we here set up a 
version of DoS that is suitable for addressing the sign problem of fermion worldlines. 

The basic idea is to define the density $\rho(n)$ as a function of the number $n$ of fermion loops with a negative sign\footnote{The fermion worldlines form 
closed loops such that from now on we will often use the term \emph{fermion loops}.}, such that the partition function $Z$ then is given as
$Z = \sum_{n=0}^\infty \rho(n) \, (-1)^n$. Two prerequisites  are needed for this idea to work: 1) An update algorithm for the fermion 
loops that allows one to control the signs of the loops and at the same time is ergodic.  2) The density $\rho(n)$ must be fast decaying with $n$, such that the 
sum for $Z$ converges quickly and can be truncated after a reasonable number of terms. In this exploratory presentation we show that for the case of staggered 
fermions indeed both prerequisites can be fulfilled, i.e., we discuss suitable Monte Carlo steps that allow to control the loop signs and present preliminary 
numerical results that illustrate (at least in 2d) exponential decay of $\rho(n)$ for large $n$, 

\section{Worldline representation for fermions and formulation of the DoS approach}

\noindent
The DoS method for fermion worldlines introduced here is a rather general 
approach, but for clarity of the presentation we discuss it for a specific model: 
The dynamical degrees of freedom are the 1-component 
Grassmann-valued fermion fields $\psi_x$ and $\overline{\psi}_x$ assigned to the 
sites $x$ of a $d$-dimensional lattice. The boundary conditions are periodic for 
$d-1$ of the dimensions and anti-periodic for dimension $d$ which is the euclidean time direction.
The corresponding action is given by
\begin{equation}
S \; =  \; \sum_{x,\nu} \gamma_{x,\nu} \, \overline{\psi}_x \frac{ e^{\, \mu \delta_{\nu,d}} \psi_{x+\hat \nu} - e^{\, - \mu \delta_{\nu,d}}\psi_{x-\hat \nu}}{2}
\; + \; M \sum_x \overline{\psi}_x \psi_x
\; - \; \frac{J}{4} \sum_{x,\nu} \overline{\psi}_x \psi_x
\overline{\psi}_{x+\hat \nu} \psi_{x+\hat \nu} \; .
\label{S_conventional}
\end{equation}
In the sums $x$ runs over the sites of the lattice and $\nu$ over the directions $1,2 \; ... \; d$, with $\hat \nu$ denoting the unit vector in direction $\nu$. 
The first sum in (\ref{S_conventional}) is the kinetic term with the staggered sign factors $\gamma_{x,\nu} = (-1)^{x_1 + x_2 \; ... \; x_{d-1}}$, the second sum the  
mass term and the last sum constitutes a quartic
self-interaction with coupling $J$.  A chemical potential
$\mu$ has been introduced by weighting the temporal ($\nu = d$) hops in the kinetic term with $e^{\pm \mu}$. 
The partition sum of the system is given by integrating the Boltzmann factor 
$e^{-S}$ with the product of Grassmann measures $\int \prod_x d\psi_x \, d \overline{\psi}_x$.

The standard approach to a Monte Carlo simulation of the system would first introduce a Hubbard Stratonovich (HS) field to 
break up the quartic interaction into a bilinear that couples to the HS field. In this form the fermions
can be integrated out giving rise to a fermion determinant. The partition sum is then obtained by integrating the fermion determinant 
over all configurations of the HS field with a Gaussian weight. However, for $\mu \neq 0$ the fermion determinant is complex such that it has
no probability interpretation and only 
the case of vanishing chemical potential is accessible for Monte Carlo simulations. While for many bosonic systems the worldline 
representation completely solves the complex action problem, this is not the case for fermions, since the fermion worldlines still have signs.
We will see, however, that the fermionic worldline picture allows for a natural implementation of a DoS approach.

\begin{figure}[t] 
\begin{center}
\vspace*{-12mm}
\includegraphics[scale=0.25,clip]{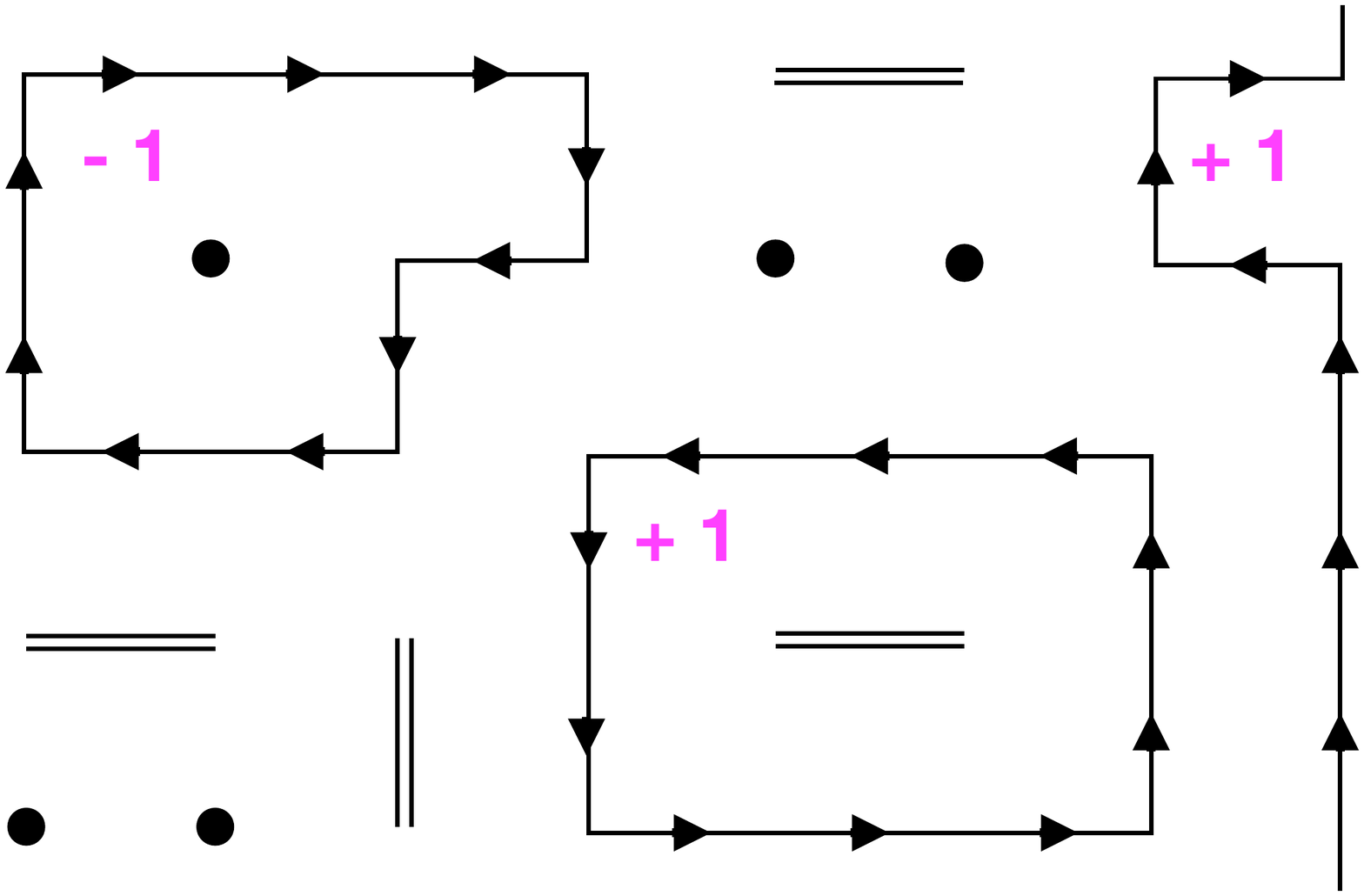}
\end{center}
\vspace*{-15mm}
\caption{Example of an admissible configuration of monomers (dots), dimers (double lines) and fermion loops (single lines with arrows) in 2d. The numbers
next to the loops give the sign of the respective loop.}
\label{figure_1}
\end{figure}

In the worldline representation the partition sum is exactly rewritten into a sum over configurations of monomers, dimers and loops. Monomers
occupy a single site, dimers two neighboring sites, i.e., a link, and the fermion loops are oriented closed contours of links. The loops may not
touch or intersect. In addition monomers, dimers and loops obey a constraint: Each site of the lattice is either occupied by a monomer, is the endpoint 
of a dimer or is run through by a loop. Fig.~\ref{figure_1} shows an example of an admissible configuration in 2d, where monomers are shown as dots,
dimers as double lines and loops as oriented single lines. The partition sum now is given by
\begin{equation}
Z \; = \; \frac{1}{2^V}\!\! \sum_{\{m,d,l\}} \! (2M)^{\#\,m} \;  (1\!+\!J)^{\# \, d} \; e^{\,\mu \beta W} \; \prod_l \mbox{sign}(l) \; = \; \frac{1}{2^V} \!\!
\sum_{\{m,d,l\}} \! (2M)^{\#\,m} \;  (1\!+\!J)^{\# \, d} \; e^{\,\mu \beta W} \; (-1)^{\, {N}} ,
\label{Z_worldline}
\end{equation}
where the sum $\sum_{\{m,d,l\}}$ over the configurations of monomers $m$, dimers $d$ and loops $l$ is a restricted sum, where only admissible 
configurations are taken into account. Each configuration comes with a weight factor where $\#\,m$ denotes the number of monomers and
$\#\,d$ the number of dimers. $\beta$ is the temporal extent of the lattice, which corresponds to the inverse temperature in lattice units and $W$ 
is the total net winding number of the loops around the compact time direction. The fact that $W$ appears as factor of $\mu \beta$ shows that 
$W$ can be identified with the net particle number. From now on we will switch to the canonical formalism, i.e., we work at a fixed net particle number,
i.e., a fixed net winding number $W$ and the factor $e^{\,\mu \beta W}$ will no longer appear.

In Eq.~(\ref{Z_worldline}) sign$(l)$ denotes the sign of a loop $l$. 
This sign has several contributions: 1) The product of staggered sign factors $\gamma_{x,\nu}$ 
along the links of the loop $l$. 2) A factor of $-1$ for every link that is run through by the loop
in negative direction. 3) An overall sign originating from the reordering of the Grassmann variables along the loop. 
4) A factor of $-1$ for every winding of the loop around 
compact time, which is due to the anti-periodic temporal boundary conditions. 
The overall sign of a configuration is then given by $\prod_l \mbox{sign}(l)$, which in the second step of (\ref{Z_worldline}) was written as $(-1)^N$,
where $N$ is the number of loops with a negative sign.

We will show below, that in the canonical setting, i.e., at fixed winding number $W$, local update steps can be combined into an ergodic algorithm. 
In each of these local update steps one can evaluate the change of the loop signs,
such that one has an ergodic algorithm with full control of the total number $N$ of loops with negative sign. Thus it makes sense to consider a density of
states that is a function of the number $n$ of negative sign loops. We define the density $\rho(n)$ as
\begin{equation}
\rho(n) \; = \; \sum_{\{m,d,l\}} (2M)^{\#\,m} \;\;  (1\!+\!J)^{\# \, d} \; \delta_{n \, , \, {N} } \; ,
\label{rhodef}
\end{equation}
where the sum over the dynamical degrees of freedom is now understood such that the net winding number $W$ remains fixed.
The Kronecker delta $ \delta_{n \, , \, {N} }$ in (\ref{rhodef}) restricts the sum over the dynamical degrees of freedom such that 
the number $N$ of negative sign loops is frozen to $N = n$. Obviously the partition sum then is given by
\begin{equation}
Z \; = \; \sum_{n = 0}^\infty  \rho(n) \; (-1)^n \; .
\label{Z_density}
\end{equation}
The inclusion of observables will be discussed in an upcoming paper and we continue with introducing a suitable parameterization for the density $\rho(n)$. 

\section{Parameterization and evaluation of the density}

\noindent
The next step is to find a suitable parameterization of the density $\rho(n)$ and a strategy for its determination.
Our parameterization is chosen piecewise constant, but it is convenient to write the constants in exponential form, 
such that the parameterization in terms of real exponents $a_n$ reads 
\begin{equation}
\rho(n) \; = \; \rho(n-1) \; e^{\, -a_n} \qquad \mbox{with} \qquad a_0 = f(M,J) \;\; \Leftrightarrow \;\; \rho(0) = e^{-f(M,\,J)} \; .
\label{density_param}
\end{equation}
The overall normalization of the density can be chosen freely with some restrictions that become important when observables 
are introduced. The normalization obviously is tied to the first exponent $a_0$ and as a normalization we choose
$a_0 = f(M,J)$ where $f(M,J)$ can be chosen as a function of the couplings. Of course all exponents $a_n$ will depend 
on the couplings implicitly, but the freedom of normalization allows for one explicit choice, i.e., our $a_0 = f(M,J)$.  

Having parameterized the density in terms of the exponents $a_n$ we now need to describe how these parameters
can be evaluated. Once they are determined one can compute the partition sum using (\ref{density_param}) and (\ref{Z_density}). 
For computing the $a_n$ we introduce a restricted partition sum defined as
\begin{equation}
Z_n (\lambda) = \! \sum_{\{m,d,l\}} \!\!\!(2M)^{\#\,m} \;  (1\!+\!J)^{\# \, d}  \;  e^{\, \lambda \, N} \;
\Theta_n(N) \!
\quad \mbox{with} \quad 
\Theta_n(N) \, = \, \left\{
\begin{array}{l}
1 \;\; \mbox{for} \; N \in \{{n}, {n+1}\}
\\
0 \;\;  \mbox{otherwise.}
\end{array}
\right.
\label{Z_restricted}
\end{equation}
The restricted partition sum $Z_n(\lambda)$ depends on the real control parameter $\lambda$ which in the sum over all configurations
couples via $e^{\, \lambda \, N}$ to the number of negative sign loops $N$. This number $N$ is restricted by the support function 
$\Theta_n(N)$ which admits only $N = n$ and $N = n+1$. We have already outlined that our updates allow one to control the 
number $N$. Furthermore no more sign factors appear in (\ref{Z_restricted}) such that $Z_n(\lambda)$ can be studied as
function of $\lambda$ using Monte Carlo simulations. 

Using the definition (\ref{rhodef}) of the density, as well as its parameterization (\ref{density_param}) we may express the restricted 
partition sum $Z_n(\lambda)$ as
\begin{equation}
Z_n(\lambda) \; = \; \rho(n) \, e^{\, \lambda \, n} \; + \;  \rho(n+1) \, e^{\, \lambda \, (n+1)} \; = \; 
\rho(n) \, e^{\, \lambda \, n} \, \left[ 1 + e^{\, \lambda - a_{n+1}} \right] \; .
\end{equation}
Taking the derivative of the logarithm of this partition sum with respect to the control parameter $\lambda$ we obtain the restricted vacuum expectation value 
of the number $N$ of negative sign loops,
\begin{equation}
\langle N \rangle_n(\lambda)  \, = \,  \frac{ \partial \! \ln \, Z_n(\lambda)}{\partial \lambda} \, = \,
\frac{1}{Z_n(\lambda)} \! \sum_{\{m,d,l\}} \!\!\!\!(2M)^{\#\,m} \; (1\!+\!J)^{\# \, d} \; \Theta_{n}(N) \; e^{\lambda \, N} \, = \, 
n \, + \, \frac{1}{2} \left[ 1 + 
\tanh \! \left(  \frac{\! \lambda \!-\! a_{n+1}}{2} \!\right) \right] ,
\end{equation}
where the third expression is the form in terms of the path integral over the dynamical variables $m, d, l$, while the final expression 
is the form based on the density. As discussed for the restricted partition sum above, also the path integral form of $\langle N \rangle_n(\lambda)$
can be evaluated with Monte Carlo techniques. After a trivial normalization we find 
$V(\lambda) \equiv 2 \langle N \rangle_n(\lambda) - 2n - 1 = \tanh \left( \frac{ \lambda -  a_{n+1}}{2} \right)$.
The Monte Carlo results for $V(\lambda)$ for different values of $\lambda$ can be fit with the simple 1-parameter function 
$\tanh ( (\lambda -  a_{n+1})/2 )$ and the exponent $a_n$ is obtained from this fit. Repeating this procedure for different $n$ we obtain 
the exponents $a_n$ and from those the density $\rho(n)$. 

\section{Monte Carlo simulation}

\noindent
We now come to presenting a set of local update steps that can be combined into an ergodic algorithm and have the property that at each update 
step we may control how the signs of the loops change. This is necessary to take into account the support function $\Theta_{n}(N)$ in the simulation 
of $\langle N \rangle_n(\lambda)$, which restricts the number $N$ of loops with negative signs to $\{n,n+1\}$. 
Of course our update  steps also have to update the monomer and dimer degrees of freedom 
and the configurations have to be admissible ones, i.e., each site has to be run through by one fermion worldline or is the endpoint of a dimer or 
is occupied by a monomer. Furthermore we need to take into account that our DoS approach uses a canonical setting, i.e., we work at a fixed temporal 
net winding number $W$ of the fermion loops. In order to implement that we use as a starting configuration a configuration with $W$ straight loops 
in direction $d$ that close around compactified time. The sites not visited by the fermion loops are then occupied with monomers. 
The update steps we discuss below 
do not change the winding number, such that the simulation remains in the sector with net particle number $W$. This also implies that we can ignore the
contribution of the anti-periodic boundary conditions to the loop signs. We finally remark that the update steps we discuss are illustrated in 2 dimensions, 
but it is straightforward to see that they work also in higher dimensions.

\begin{figure}[t] 
\begin{center}
\hspace*{-1mm}
\includegraphics[scale=0.32,clip]{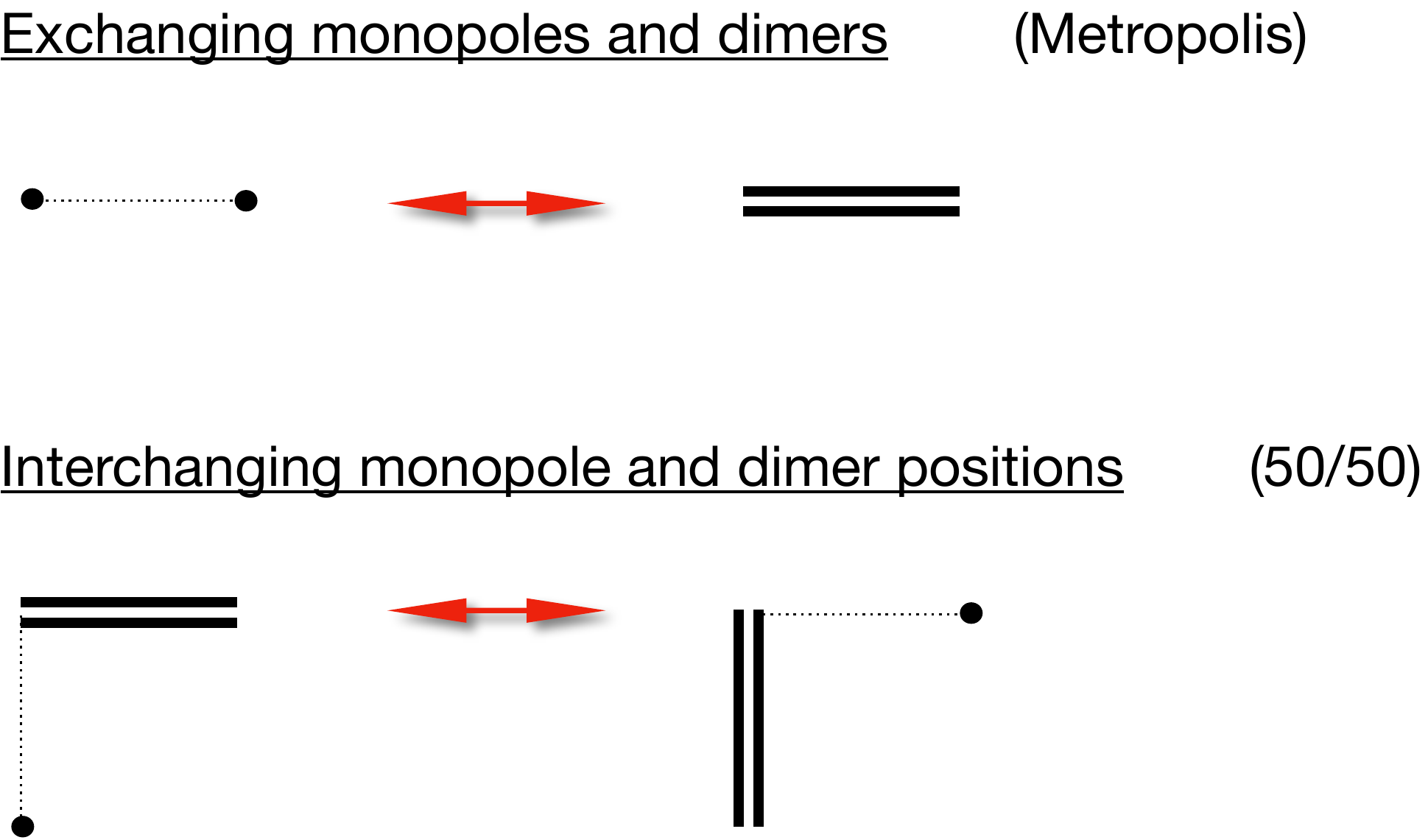}
\hspace{6.5mm}
\includegraphics[scale=0.32,clip]{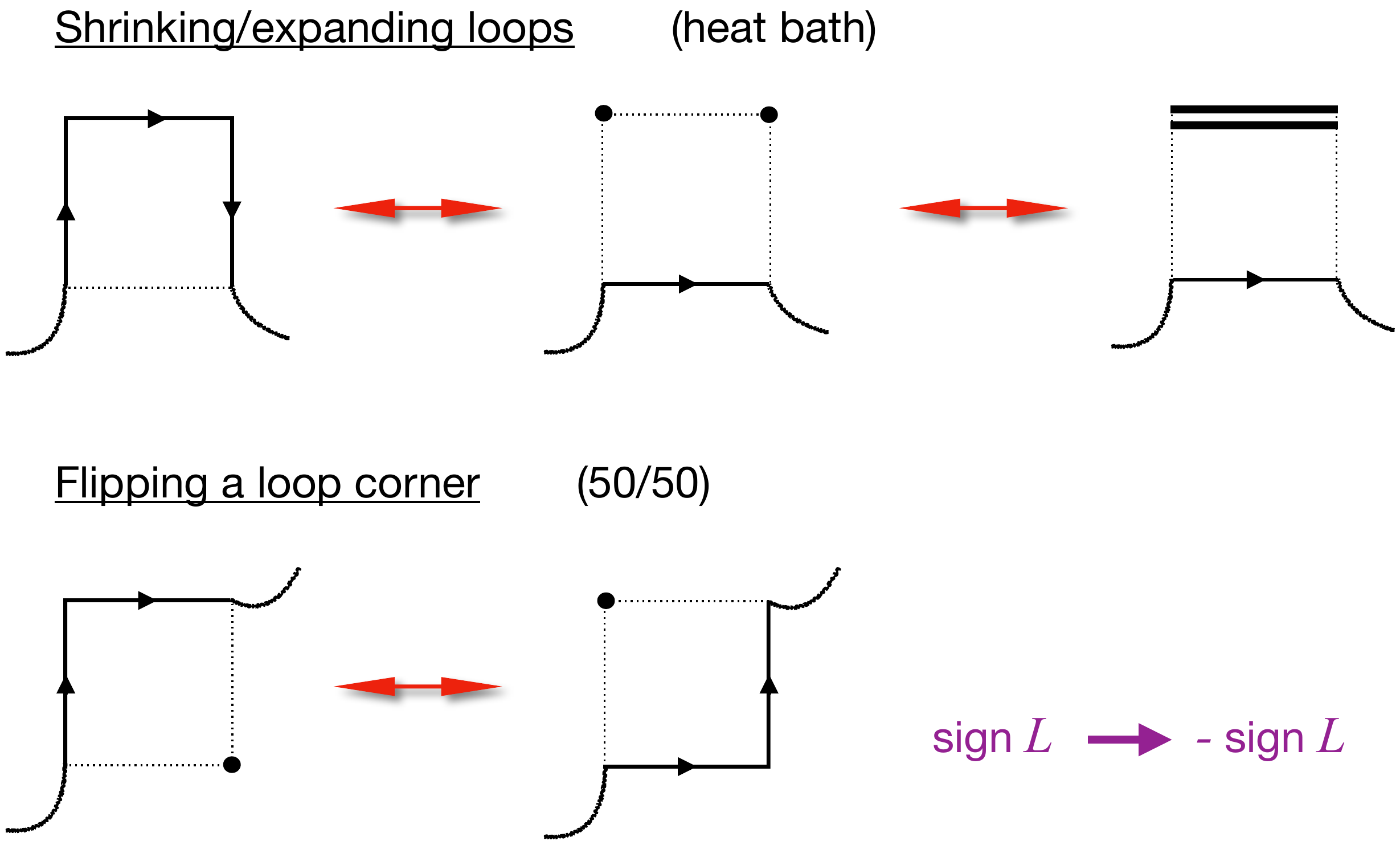}
\end{center}
\vspace*{-5mm}
\caption{Elementary update steps of our algorithm. Top left: Exchange of two monomers on neighboring sites with the corresponding dimer (accepted with 
a Metropolis step). Bottom left: Shift of a dimer and a monomer on an adjacent site (accepted with probability 1/2). Top right: Expanding or shrinking a loop 
around three sides of a plaquette with a heat bath step. The sign of the loop does not change. Bottom right: Flipping the corner of a loop. This step is accepted with
probability 1/2 and changes the sign of the loop.}
\label{figure_2}
\end{figure}

We begin the discussion of the updates with two steps that only involve monomers and dimers. They are depicted in the lhs.\ plots of Fig.~\ref{figure_2}. Obviously
an admissible configuration remains admissible if we exchange a dimer on some link by two monomers at the endpoints of the link. We accept such a step with 
the corresponding Metropolis probability (Fig.~\ref{figure_2}, lhs., top). Another step we use is to identify a dimer with a monomer on a site neighboring one of 
its endpoints and then shift the dimer towards the monopole site and place the monomer at the now empty site (Fig.~\ref{figure_2}, lhs., bottom). Here the 
weights are not changed and we accept this move with probability 1/2. We remark that this step is equivalent to two of the previous steps, which, however, 
might have a poor acceptance probability at small mass.  

Next we come to a set of update steps that change the contour of a loop. The first one (top right of Fig.~\ref{figure_2}) changes a loop along three sides of
a plaquette by either shrinking or expanding the loop around that plaquette and adding or removing either two monomers or a dimer. It is straightforward to
show that the sign of the loop remains the same, since a minus sign from adding/removing a flux in negative direction is compensated by a
minus sign from the staggered sign factors. The change of weight from adding/removing dimers or monomers is taken into account by a heat bath.

Another way of altering the loop is shown in the bottom right of Fig.~\ref{figure_2}, where we flip the corner of a loop and move one monomer.
Here the weight does not change, such that this step is accepted with probability 1/2. Here, however, the sign of the loop is flipped, due to the change
of the staggered sign factors. This implies that this update step can be admitted only if also after the step the number $N$ of negative loops is still
in $\{n,n+1\}$ as required by the support function $\Theta_n(N)$.

\begin{figure}[t] 
\begin{center}
\includegraphics[scale=0.35,clip]{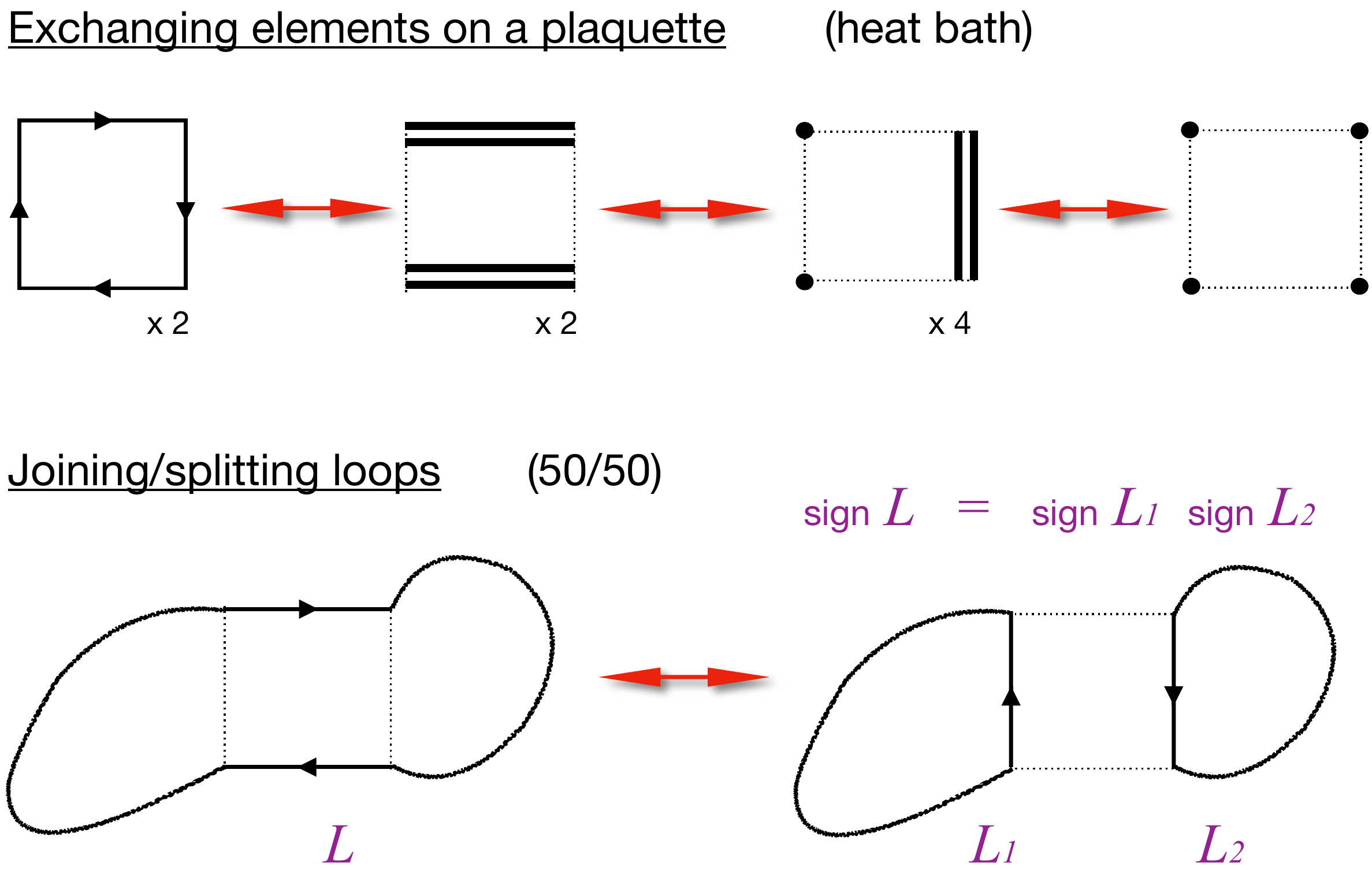}
\end{center}
\vspace*{-5mm}
\caption{Update steps that change the number of loops. Top: Combinations of dimers and monomers that fill the sites and links of a 
plaquette can be exchanged with a loop around the plaquette (heat bath). Bottom: Changing two anti-parallel fluxes on a plaquette splits 
a single loop $L$ into two loops $L_1, L_2$ or vice versa. The step is accepted with probability 1/2 and the loop signs obey sign $L$ = 
sign $L_1$ sign $L_2$.}
\label{figure_3}
\end{figure}

Finally we come to update steps that change the number of loops. The first one, illustrated in the top of Fig.~\ref{figure_3} inserts/removes an
elementary loop around a single plaquette by removing/inserting a combination of dimers or monomers or both. The elementary loop has positive sign,
and when inserting such a loop the orientation can be chosen randomly. The change of weight from adding/removing the dimers or
monomers can again be taken into account by a heat bath.

The last step we discuss, is a re-routing of two antiparallel fluxes on opposite links of a plaquette. As the bottom plot of Fig.~\ref{figure_3} shows,
this either gives rise to splitting a single loop $L$ into two loops $L_1$ and $L_2$ or the inverse step by fusing two loops $L_1$ and $L_2$
into a single loop $L$. One finds that the signs of the loops obey sign$(L) = $ sign$(L_1) $ sign$(L_2)$, which can be seen from the property that the
product of the staggered sign factors over the links of a plaquette equals $-1$ and the fact that each individual loop has an overall factor of
$-1$. The weights do not change here, such that the acceptance probability is 1/2, but of course the number of negative loops can change,
implying that also this step can only be offered if the number of negative loops remains in $\{n,n+1\}$.

\begin{figure}[t] 
\begin{center}
\hspace*{-5mm}
\includegraphics[height=60mm,clip]{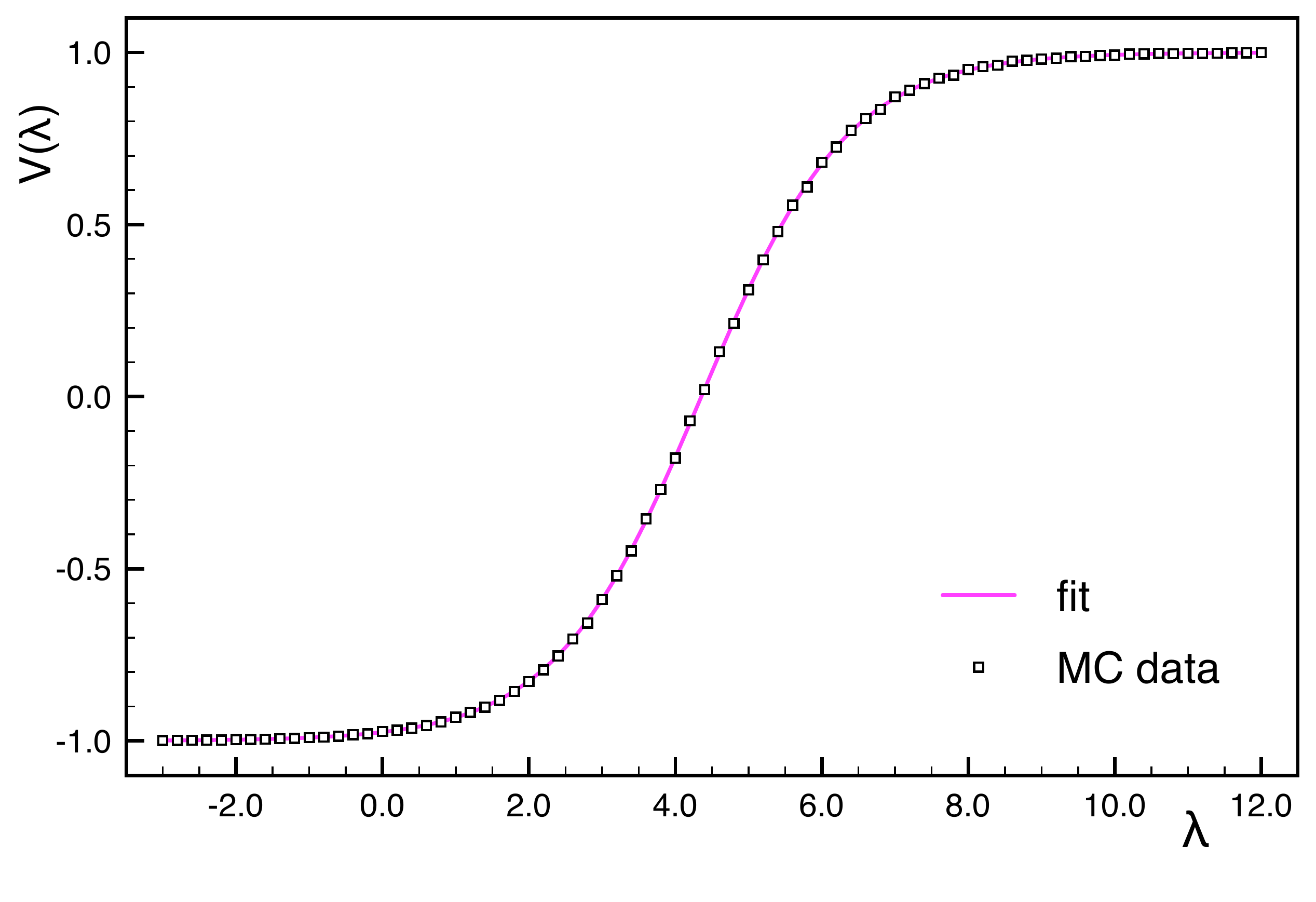}
\includegraphics[height=60mm,clip]{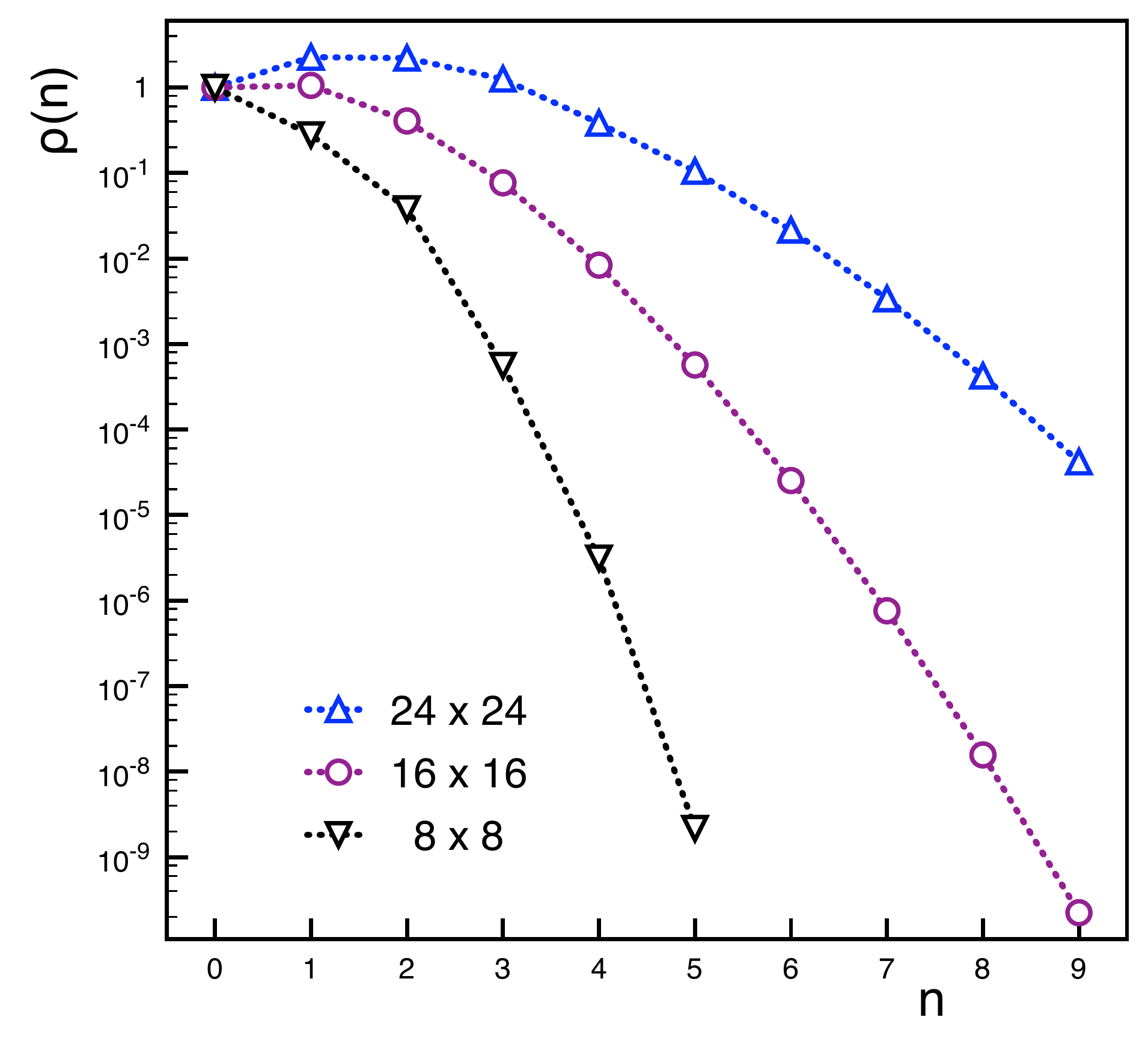}
\end{center}
\vspace*{-5mm}
\caption{Left: Example for the fit of the data for $V(\lambda)$ with $\tanh((\lambda - a_{n+1})/2)$. Right: The density $\rho(n)$ 
as a function of $n$ for different lattice volumes.}
\label{figure_4}
\end{figure}

For a first test of the new approach we set up a simulation of our model in 2 dimensions.  We work on lattices of size $L \times L$ and consider $L = 8, 16, 24$ at vanishing 
winding number, i.e., we set $W = 0$. The different update steps of our simulation strategy discussed in the previous paragraphs are combined into sweeps where a full 
sweep is defined as applying each update step once to all links or plaquettes it can act on. For our simulations we typically use $10^5$ sweeps for equilibration 
and a statistics of $10^5$ measurements separated by 10 sweeps for decorrelation. We determine $\langle N \rangle_n(\lambda)$ for typically 50 values
of $\lambda$ and fit the corresponding $V(\lambda)$ with $\tanh((\lambda - a_{n+1})/2)$ to determine the exponent $a_{n+1}$. An example of this step 
is shown in the lhs.\ plot of Fig.~\ref{figure_4}. Obviously the data (symbols) are described by the fit function (continuous curve) extremely well. 

From the exponents $a_n$ one can determine the density $\rho(n)$ using Eq.~(\ref{density_param}). To get a first impression of its properties,
in the rhs.\ plot of Fig.~\ref{figure_4} we show $\rho(n)$ as a function of $n$ for different volumes, using a normalization 
such that $\rho(0) = 1$, i.e., $a_0 = 0$. The plot is for couplings $M = 0.5$ and $J = 0.0$. Here the main question is to understand 
whether the density is fast decreasing for large $n$. Obviously  
this is the case (note the logarithmic scale on the vertical axis). However, it is obvious that for the larger two volumes a maximum of $\rho(n)$ 
appears for some $n_{max} > 0$, which implies that for increasing volume the most likely number of negative sign loops is at this 
$n_{max} > 0$. Only  for $n > n_{max}$ the exponential decrease of the density sets in. One naturally expects that a negative sign loop has a characteristic 
coupling-dependent size\footnote{Note that the smallest possible negative sign loop is around a $3 \times 3$ square.}, such that with increasing volume more negative sign
loops of the characteristic size fit on the lattice. Thus one expects, that in leading order $n_{max}$ scales proportional to the volume, such that also the range of values $n$ where
$\rho(n)$ needs to be evaluated until $\rho(n)$ becomes sufficiently small scales with the volume. This expectation that the range where the density has to be computed 
scales with the volume is in agreement with the experience from other DoS applications. 

The first numerical tests presented here show that indeed the density $\rho(n)$ is fast decreasing for sufficiently large $n$, and it is plausible that the range of 
$n$ where $\rho(n)$ needs to be evaluated scales linearly with the system size. Currently we are working on the implementation of observables and prepare a systematic 
comparison of the DoS results for free fermions to the corresponding exact calculation, in order to assess whether the necessary accuracy of $\rho(n)$ can be achieved, such that 
an efficient use of the new method is feasible. 

\begin{thebibliography}{99}

\bibitem{Endres:2006zh}
M.~G.~Endres,
PoS \textbf{LAT2006}, 133 (2006)
doi:10.22323/1.032.0133
[arXiv:hep-lat/0609037]

\bibitem{Endres:2006xu}
M.~G.~Endres,
Phys. Rev. D \textbf{75}, 065012 (2007)
doi:10.1103/PhysRevD.75.065012
[arXiv:hep-lat/0610029]

\bibitem{Wolff:2010qz}
U.~Wolff,
Nucl. Phys. B \textbf{832}, 520-537 (2010)
doi:10.1016/j.nuclphysb.2010.02.005
[arXiv:1001.2231]

\bibitem{Korzec:2010sh}
T.~Korzec, U.~Wolff,
PoS \textbf{LATTICE2010}, 029 (2010)
doi:10.22323/1.105.0029
[arXiv:1011.1359]

\bibitem{Gattringer:2012df}
C.~Gattringer, T.~Kloiber,
Nucl. Phys. B \textbf{869}, 56-73 (2013)
doi:10.1016/j.nuclphysb.2012.12.005
[arXiv:1206.2954]

\bibitem{DelgadoMercado:2013ybm}
Y.~Delgado Mercado, C.~Gattringer, A.~Schmidt,
Phys. Rev. Lett. \textbf{111} 141601 (2013)
doi:10.1103/PhysRevLett.111.141601
[arXiv:1307.6120]

\bibitem{Gattringer:2014nxa}
C.~Gattringer,
PoS \textbf{LATTICE2013}, 002 (2014)
doi:10.22323/1.187.0002
[arXiv:1401.7788]

\bibitem{Gattringer:2016lml}
C.~Gattringer, C.~Marchis,
Nucl. Phys. B \textbf{916}, 627-646 (2017)
doi:10.1016/j.nuclphysb.2017.01.025
[arXiv:1609.00124]

\bibitem{Marchis:2017oqi}
C.~Marchis, C.~Gattringer,
Phys. Rev. D \textbf{97}, 034508 (2018)
doi:10.1103/PhysRevD.97.034508
[arXiv:1712.07546]

\bibitem{Gattringer:2018ous}
C.~Gattringer, M.~Giuliani, O.~Orasch,
Phys. Rev. Lett. \textbf{120}, no.24, 241601 (2018)
doi:10.1103/PhysRevLett.120.241601
[arXiv:1804.01580]

\bibitem{Gattringer:2018dlw}
C.~Gattringer, D.~G\"oschl, T.~Sulejmanpasic,
Nucl. Phys. B \textbf{935}, 344-364 (2018)
doi:10.1016/j.nuclphysb.2018.08.017
[arXiv:1807.07793]

\bibitem{Sulejmanpasic:2019ytl}
T.~Sulejmanpasic, C.~Gattringer,
Nucl. Phys. B \textbf{943}, 114616 (2019)
doi:10.1016/j.nuclphysb.2019.114616
[arXiv:1901.02637]

\bibitem{Sulejmanpasic:2020lyq}
T.~Sulejmanpasic, D.~G\"oschl, C.~Gattringer,
Phys. Rev. Lett. \textbf{125}, 201602 (2020)
doi:10.1103/PhysRevLett.125.201602
[arXiv:2007.06323]

\bibitem{Anosova:2022cjm}
M.~Anosova, C.~Gattringer,  T.~Sulejmanpasic,
JHEP \textbf{04}, 120 (2022)
doi:10.1007/JHEP04(2022)120
[arXiv:2201.09468]

\bibitem{Gattringer:2015nea}
C.~Gattringer, T.~Kloiber, V.~Sazonov,
Nucl. Phys. B \textbf{897}, 732-748 (2015)
doi:10.1016/j.nuclphysb.2015.06.017
[arXiv:1502.05479]

\bibitem{Goschl:2017kml}
D.~G\"oschl, C.~Gattringer, A.~Lehmann, C.~Weis,
Nucl. Phys. B \textbf{924}, 63-85 (2017)
doi:10.1016/j.nuclphysb.2017.09.006
[arXiv:1708.00649]

\bibitem{Hirtler:2022ycl}
D.~Hirtler, C.~Gattringer, contribution to PoS {\bf LATTICE 2022},
[arXiv:2210.13787]

\bibitem{Francesconi:2021suq}
O.~Francesconi, {\it The sign problem in particle systems}, PhD Thesis, Swansea University 2021.

\bibitem{Gocksch1}
A.~Gocksch, P.~Rossi, U.M.~Heller,
Phys.\ Lett.\ B {\bf 205} (1988) 334,
doi:10.1016/0370-2693(88)91674-7

\bibitem{Gocksch2}
A.~Gocksch,
Phys.\ Rev.\ Lett.\  {\bf 61} (1988) 2054,
doi:10.1103/PhysRevLett.61.2054

\bibitem{Langfeld:2012ah}
  K.~Langfeld, B.~Lucini, A.~Rago,
  Phys.\ Rev.\ Lett.\  {\bf 109} (2012) 111601,
  doi:10.1103/PhysRevLett.109.111601,
    [arXiv:1204.3243]

\bibitem{Langfeld:2013xbf}
  K.~Langfeld, J.M.~Pawlowski,
  Phys.\ Rev.\ D {\bf 88} (2013) 071502,
  doi:10.1103/PhysRevD.88.071502,
    [arXiv:1307.0455]
 
\bibitem{Langfeld:2014nta}
  K.~Langfeld, B.~Lucini,
  Phys.\ Rev.\ D {\bf 90} (2014) 094502,
  doi:10.1103/PhysRevD.90.094502,
    [arXiv:1404.7187]

\bibitem{Gattringer:2016kco}
  C.~Gattringer, K.~Langfeld,
 Int.\ J.\ Mod.\ Phys.\ A {\bf 31} (2016)  1643007,
 doi:10.1142/S0217751X16430077,
 [arXiv:1603.09517]

\bibitem{Garron:2016noc}
  N.~Garron, K.~Langfeld,
  Eur.\ Phys.\ J.\ C {\bf 76} (2016)  569,
  doi:10.1140/epjc/s10052-016-4412-2,
    [arXiv:1605.02709]

\bibitem{Garron:2017fta}
  N.~Garron, K.~Langfeld,
  Eur.\ Phys.\ J.\ C {\bf 77} (2017)  470,
  doi:10.1140/epjc/s10052-017-5039-7,
  [arXiv:1703.04649]
  
\bibitem{Francesconi:2019nph} 
O.~Francesconi, M.~Holzmann, B.~Lucini, A.~Rago,
Phys. Rev. D \textbf{101}, 014504 (2020),
doi:10.1103/PhysRevD.101.014504,
[arXiv:1910.11026]  

\bibitem{Z3_FFA_2}
  C.~Gattringer, P.~T\"orek,
  Phys.\ Lett.\ B {\bf 747} (2015) 545,
  doi:10.1016/j.physletb.2015.06.017,
  [arXiv:1503.04947]
  
\bibitem{FFA_2}
  M.~Giuliani, C.~Gattringer, P.~T\"orek,
  Nucl.\ Phys.\ B {\bf 913} (2016) 627,
  doi:10.1016/j.nuclphysb.2016.10.005,
  [arXiv:1607.07340]
  
  \bibitem{FFA_3}
  M.~Giuliani, C.~Gattringer,
  Phys.\ Lett.\ B {\bf 773} (2017) 166,
  doi:10.1016/j.physletb.2017.08.014,
  [arXiv:1703.03614]
  
\bibitem{Gattringer:2019khb}
  C.~Gattringer, M.~Mandl, P.~Törek,
  Phys.\ Rev.\ D {\bf 100} (2019) 114517,
  doi:10.1103/PhysRevD.100.114517,
  [arXiv:1911.05320]

\bibitem{Gattringer:2019egx}
  C.~Gattringer, M.~Mandl, P.~T\"orek,
  Particles {\bf 2020 03} (2020),
  doi:10.3390/particles3010008,
  [arXiv:1912.05040]
  
\bibitem{Gattringer:2020mbf}
C.~Gattringer, O.~Orasch,
Nucl. Phys. B \textbf{957} (2020), 115097,
doi:10.1016/j.nuclphysb.2020.115097,
[arXiv:2004.03837]

\bibitem{Gattringer:2021xrb}
C.~Gattringer, O.~Orasch,
PoS \textbf{LATTICE2021}, 158 (2022)
doi:10.22323/1.396.0158,
[arXiv:2111.09535]


\end{thebibliography}
\end{document}